\documentclass[12pt,aps,showpacs,amsmath,amssymb]{revtex4}

\usepackage{graphicx}
\usepackage{dcolumn}

\usepackage{color}
\newfont{\largemi}{cmmi10}

\newfont{\smallmi}{cmmi6}

\begin{document}

\title{Quantum Statistical Transition }
\author{Tao Wang}
\email{taowang@thnu.edu.cn}
\affiliation{Department of Physics,
Tonghua Normal University, Tonghua, 134002, P.R.China}

\begin{abstract}
By analyzing the BCS-BEC crossover, I found that because of the
pairing interactions, a continuous family of quantum statistics
interpolating between fermions and bosons is possible, although it
seems incapable to construct reasonable wave function.
\end{abstract}


\date{\today}
\maketitle

It is known that particles can be divided into two categories:
fermions and bosons. Fermions have half-integer spin, while bosons
have zero or integer spin. Composite particles, such as atoms, also
fall into one of the two, depending on the total spin of their
constituent particles.

At the quantum level, fermions  can not occupy the same quantum
state at the same time and obey Fermi-Dirac statistics (FDs), while
the bosons can share quantum states and satisfy Boson-Einstein
statistics (BEs). At sufficiently low temperatures, fermions tend to
fill energy states from the lowest up, with one particle per quantum
state. The quantum behavior emerges gradually as the fermions gas is
cooled below the Fermi temperature $T_{F}=\varepsilon_{F}/k_{B}$,
where $\varepsilon_{F}$ is the Fermi energy and $k_{B}$ is
Boltzmann' s constant. $T_{F}$¬¬ marks the crossover from the
classical to the quantum regime. Whereas bosons will rapidly
collapse into an individual quantum state to form a Boson-Einstein
condensate, and the transition temperature $T_{c}$¬¬ is a
determinate point.

Each quantum system, the fermion system or boson system, has been
investigated for a long time, but research on the relationship
between them is at the beginning. Although the process for a single
boson converting automatically into an individual fermion or the
opposite can not be observed, the continuous transition from a Fermi
system to a Boson system or vice versa has been realized which is
called BCS-BEC crossover~\cite{BCS,BEC,crossover}, and the key
factor is the pairing up of two fermions.

Pairing correlations play a crucial role in many Fermi systems. If
the attractive interaction between two Fermions is weak, the pairing
correlation can be understood in terms of the
Bardeen-Cooper-Schrieffer(BCS) theory, which show a strong
correlation in the momentum space. If the interaction is
sufficiently strong, on the other hand, two fermions will form a
bosonic bound state and condense in the ground state of a many-body
system. The transition from the BCS-type pairing correlations to the
BEC-type takes place continuously as a function of the strength of
the pairing interaction.

What I will discuss in this short paper is not how the BCS-BEC
crossover evolves but whether there is something new in the
transition which results from pairing interactions. The reasons I
propose this question are as followings: (1) Cooper-pairs are not
real bosons which can only be produced  closed to the Fermi surface
and have both bosonic and fermion properties. For the BCS-BEC
crossover, what usually considered is how the pairing correlations
occurs  change, but the way by which the fermion property vanishes
is ignored for all the time; (2) Fermi and Boson systems have
distinct characteristics. The spins of the two systems are two
different numbers, but the transition between them is continuous, an
apparent paradox.

The conflict will be more evident in the following ideal process.
 When pairing interaction is gradually added to an ideal Fermi
gas, the ideal process from an ideal gas of fermions to an ideal
Boson gas is naturally given. In the process of transition, the
interaction only bears on two certain fermions, so the pairs of two
fermions have no real force mutually and the correlations among them
can be only statistical, which originate from the competition
between the FDs and BEs.   If the strength of interaction is the
same for all pairs, the pairs are also identical for one another.
When adding weak pairing interactions, bosonic behavior will emerge
and exist together with fermion property. The former plays a key
role in the BCS mechanism, and the latter still keep the system
 as a Fermion one.  While on the side of Boson system, there is only bosonic
behavior. Therefore, in the ideal process two transitions coexist
from Cooper pairs to bosons and from a Fermi system to a Boson one,
and the latter must have an important impact on the collective
behavior of the system, especially the crossover regime.

In the ideal process, if the BCS effect is not considered, the
Fermion system will spontaneously and continuously convert into a
Boson system, and the fermion and bosonic properties act on the same
pairs simultaneously. Therefore, the statistics in the transition is
neither FDs nor BEs, and a continuous transition between the FDs and
BEs is possible, which is called quantum statistical transition(QST)
and here this kind of statistics is named as transitional
statistics(Ts).

 To realize the Ts mathematically, a technical hypothesis
should be offered, in which the FDs and BEs can be unified into a
single expression and become two limiting cases of the single
expression. Considering the different ways of occupation on a single
quantum state, 1 for Fermions and $\infty$ for Bosons, a possibility
can be given here: in the process of transition, the occupation in
one individual state can be any value, from 1 to $\infty$. In this
way two conclusions can be obtained from the hypothesis: (1) At zero
temperature the occupation volume of particles in the energy states
space will gradually collapse into a point; (2) For general
situation, the corresponding contribution for the maximal occupation
number $\eta$  can be easily derived from the theory of grand
canonical ensembles, summing from 0 to $\eta$ ,which is
\begin{equation}
a_{l}=\frac{\omega_{l}}{e^{\frac{\varepsilon_{l}-\mu}{k_{B}T}}-1}-
\frac{(\eta+1)\omega_{l}}{e^{(\eta+1)\frac{\varepsilon_{l}-\mu}{k_{B}
T}}-1}
\end{equation}
where $\varepsilon_{l}(l=1,2,\cdots)$,  $\omega_{l}$ and
 $a_{l}$  is the energy levels of a single
particle,  the degree of degeneracy for $\varepsilon_{l}$ and
 the average occupation numbers for $\varepsilon_{l}$,respectively. It
is easy to see that the expression will be reduced to the FDs when
$\eta=1$ and BEs for $\eta=\infty$. The chemical potential $\mu$ is
determined by
\begin{equation}
\sum\limits_{l}a_{l}=N.\label{1.0}
\end{equation}
where $N$ is effective total number of particles in the system.

 As an example, a system with free particle in three dimension
is studied. At zero temperature, the Fermi sphere will shrink into a
point and some concepts, such as Fermi surface, Fermi energy and
Fermi temperature should be generalized correspondingly, and they
are all functions of parameter $\eta$ . Thus in the process of the
transition, the equation is
\begin{equation}
\frac{2g\pi}{h^{3}}(2m)^{3/2}\eta\frac{2}{3}\varepsilon_{F}^{3/2}=n,\\
\varepsilon_{F}=k_{B}T_{F},\label{1.1}
\end{equation}
Then
\begin{equation}
\varepsilon_{F}=\frac{\hbar^{2}}{2m}\left(\frac{6\pi^{2}n}{g\eta}\right)^{2/3}.
\end{equation}
where $g$ is other degrees of freedom, such as spin, $m$ is
effective mass, $n$ is effective
 particle density, which are also functions of parameter $\eta$. When  $T>0$,
Eq. (\ref{1.0}) becomes
 \begin{equation}
\frac{2g\pi}{h^{3}}(2m)^{\frac{3}{2}}\int^{\infty}_{0}\left(
\frac{1}{e^{\frac{\varepsilon-\mu}{k_{B}T}}-1}-
\frac{\eta+1}{e^{(\eta+1)\frac{\varepsilon-\mu}{k_{B}
T}}-1}\right)\varepsilon^{\frac{1}{2}}d\varepsilon=n.\label{1.2}
\end{equation}
 If $\mu=0$ , the corresponding temperature $T_{0}$ can be derived, which is
 \begin{equation}
T_{0}=\frac{2\pi\hbar^{2}}{(2.612)^{2/3}mk_{B}}\frac{(n/g)^{2/3}}{(1-\frac{1}{\sqrt{\eta+1
}})}=\frac{0.639\eta^{2/3}T_{F}}{\pi^{1/3}(1-\frac{1}{\sqrt{\eta+1
}})}.
\end{equation}

It is known that when the temperature decrease gradually, the
chemical potential $\mu$ will increase. But $\mu$  has a maximum
value - the Fermi energy $\varepsilon_{F}$ - and the temperature at
$\varepsilon_{F}$  is $T_{c}$. When the temperature is chilled below
$T_{c}$, the $\mu$ remains unchanged. The key result is that an
universality exists. Defining a dimensionless quantity
$y=\varepsilon_{F}/(k_{B}T_{c})$,  from Eq. (\ref{1.1}) and Eq.
(\ref{1.2}) an equation can be gained, that is
\begin{equation}
\int^{\infty}_{0}\left( \frac{1}{e^{x-y}-1}-
\frac{\eta+1}{e^{(\eta+1)(x-y)}-1}\right)x^{1/2}dx=\frac{2}{3}y^{3/2}\eta,
\end{equation}
Therefore, a function $y$ of parameter $\eta$  can be obtained, but
one can be seen easily that it is not analytical for general case.
The behavior of the QST is decided by the dimension of system and
the number of micro-states of particles. For a Boson gas, $T_{c}$ is
the temperature for Boson-Einstein condensation, and $y=0$. For a
gas of fermions, $T_{c}=0$ , $y=\infty$. For the ideal process
$T_{c}<T_{0}$ . The precise result is not clear and must be
calculated by numerical method.

 In summary, the idea of the QST and logic for realization of
it in BCS-BEC crossover was discussed. Because of the effect of BEC,
the ideal process can not be performed in practice, but it is
physical and reasonable.
The Ts originates from the competition between the FDs and BEs , and
it is independent of certain interactions. The method of realization
for the QST is also given, which can not be fully accepted in the
framework of quantum mechanics since it is unable to construct
reasonable wave function - in fact how to gain bosonic wave function
from a fermion one is a difficult work for all the time in many
fields, such as the interacting boson model in nuclear
physics~\cite{IBM}.  Therefore, it should be a new phenomenon which
has not been recognized by the physical community. Otherwise the Ts
can be realized in any dimensional space and it is different from
fractional statistics~\cite{fraction}  which can only come true in
two dimensional space for its special topological property. The QST
will play a vital role in understanding the BCS-BEC crossover
completely and may influence the properties of the other Fermi
systems with strong correlation, such as atomic nuclei and
high-temperature superconductivity.

\vskip 0.50cm

This work was supported by the THNU research program and I thank
Yan-an Luo for his useful help.


\end{document}